\documentclass[showpacs,amssymb,10pt,reprint,aps,prd,longbibliography,nofootinbib,floatfix]{revtex4-2}
\usepackage{graphicx,epsfig,amssymb} 
\usepackage{amsmath,amsfonts, times}
\usepackage{bm} 
\usepackage{epstopdf}
\usepackage[linktocpage,colorlinks]{hyperref}
\usepackage[caption=false]{subfig}
\usepackage[usenames]{color}     
\usepackage{natbib}
\usepackage{soul}
\usepackage[utf8x]{inputenc}
\usepackage[usenames]{color}
\usepackage{float}
\definecolor{darkgreen}{rgb}{0,0.5,0}

\def\p{\partial}
\newcommand{\sss}{\scriptscriptstyle}

\def\Q{\text{\textit{Q}}}

\begin{document}
	\title{Absorption by stringy black holes}

	\author{S\'ergio V. M. C. B. Xavier}
	\email{sergio.xavier@icen.ufpa.br}
	\author{Lu\'is C. B. Crispino}
	\email{crispino@ufpa.br}
	\affiliation{Programa de P\'os-Gradua\c{c}\~{a}o em F\'{\i}sica, Universidade 
		Federal do Par\'a, 66075-110, Bel\'em, Par\'a, Brazil.}
	
	\author{Carolina L. Benone}
	\email{benone@ufpa.br} 
	\affiliation{Campus Universitário Salinópolis,
		Universidade Federal do Par\'a, 68721-000, Salinópolis, Par\'a, Brasil}

	\begin{abstract}
	We investigate the absorption of planar massless scalar waves by a charged rotating stringy black hole, namely a Kerr--Sen black hole. We compute numerically the absorption cross section and compare our results with those of the Kerr-Newman black hole, a classical general relativity solution. In order to better compare both charged black holes, we define the ratio of the black hole charge to the extreme charge as $Q$. We conclude that Kerr--Sen and Kerr-Newman black holes have a similar absorption cross section, with the difference increasing for higher values of $Q$.	
	\end{abstract}
	
	\date{\today}
	
	\maketitle

	\section{Introduction}\label{sec:int}
	
	The validity of general relativity (GR) in the regime of weak gravitational fields is unquestionable, since it was extensively tested and verified~\cite{Will:2006,CP:2020,CK:2019}. In the last decade, following suitable technological advancements,  tests of gravity in the strong field regime became more recurrent. Observations such as the first detection of gravitational waves (GW) by the LIGO and VIRGO collaborations~\cite{LIGO:2015} and the recent results of the Event Horizon Telescope (EHT) collaboration~\cite{M87_1:2019} made an object stand out as the protagonist of these experiments: the black hole. Emerged in literature as solutions of Einstein's GR equations, black holes are astrophysical objects that have an event horizon.
	
		Studies about black holes started with the first exact solution of Einstein's field equations, namely the Schwarzschild solution \cite{Schwarzschild:1916}, which describes a spherically symmetric black hole parametrized only by its mass.  Over the years, other solutions emerged, as a charged \cite{Reissner:1916,Nordstrom:1918}, a rotating \cite{Kerr:1963} and a charged and rotating generalization of the Schwarzschild solution \cite{Newman_etal:1965}. Despite appearing for the first time in the context of GR, black holes are not exclusive predictions of this theory. Other theories of gravity also present such astrophysical objects as possible solutions to their field equations, making black holes optimal candidates for investigating which gravitational theory is realized in nature.   
	
	 	On the other hand, the search for a quantum gravity theory has also increased along the years. In this context, one of the candidates for such theory is the heterotic string theory. In the low-energy limit of this model of gravity, there  exists a charged rotating black hole solution, namely the Kerr-Sen (KS) black hole \cite{Sen:1992}, which has been attracting a lot of attention.  Several studies on the features of KS black holes were made, including shadow \cite{Konoplya_etal:2016,Xavier:2020egv}, Hawking radiation \cite{Khani:2013}, absorption of charged particles and clouds \cite{Bernard:2017,Bernard:2016}, hidden symmetries \cite{Hioki_Miyamoto:2008}, cosmic censorship \cite{Gwak:2017,Siahaan:2016} and merger estimates \cite{Siahaan:2020}.
	 	
	 	Although it is expected that black holes in  astrophysical scenarios do not have significant electric charge, 		in certain contexts it is reasonable to consider charged black holes \cite{Cardoso_etal:2016}.  Besides  that, recent studies, as the one fulfilled by Zhang \cite{Zhang:2016}, point out that the assumption of a charged rotating black hole could explain some astrophysical phenomena.
	 	
	 	Nowadays, the major branches of experimental search for black holes are GW astronomy, shadow images and x-ray spectroscopy. All of these three branches depend, in a sense, on the effect of the black hole in its environment. This motivates the study of absorption and scattering of waves and particles by black holes. There is a vast literature about this topic in many different black hole scenarios (cf., e.~g.,  \cite{Dolan:2008,OCH2011,CDHO2014,CDHO2015} and references therein).

		Scalar fields stand out as the simplest proposed candidates to explain some issues in physics,  as, for instance, the abundance of dark matter \cite{Bertone_Tait:2018} and the strong-CP problem~\cite{Peccei_Quinn:1977,Peccei_Quinn:1977PRD}. Related to the latter, some attempts suggested the axion hypothesis to elucidate the recent reported results of North American Nanohertz Observatory for Gravitational Waves (NANOGrav) collaboration \cite{NANOGrav:2020,Ramberg_Visinelli:2021}. Moreover, scalar waves have been extensively studied in the investigation of absorption  and scattering by black holes.~\cite{Caio_Crispino:2013,L_C_C:2017,Haroldo_etal:2020}
		
		We address the problem of the absorption of fields by a charged rotating black hole in the context of heterotic string theory. We focus on analyzing the absorption cross section of planar massless scalar waves by a KS black hole. Due to the similarity between KS black holes and Kerr-Newman (KN) black holes, the charged rotating black hole solution of GR, we shall compare our results with those of Ref.~\cite{L_C_C:2017}. 
		
		The remainder of this paper is organized as follows. In Sec.~\ref{sec:1} we review the spacetimes of the charged rotating black holes treated in this paper, namely, KS and KN black holes. In Sec.~\ref{sec:2}  we study the dynamics of a massless scalar field impinging upon a KS black hole. In Sec.~\ref{sec:3} we use the partial wave analysis to express the absorption cross section and investigate the low and high-frequency regime. In Sec.~\ref{sec:4} we exhibit a selection of our numerical results of the absorption cross section for different values of the black hole charge and spin. We compare our results with those obtained in the KN case. Our final remarks are presented in Sec.~\ref{sec:remarks}. Throughout this paper we use natural units ($c=G=\hbar=1$) and the metric signature ($+,-,-,-$).

\section{Charged rotating black holes} \label{sec:1}
The KS black hole is a solution of the low-energy limit of the heterotic string theory. The action that describes this theory in four dimensions, in the Einstein frame, is the following \cite{Delgado_etal:2016}:
	\begin{equation}
		\begin{split}
			S = \int d^{4}x\sqrt{-g}\left(R-e^{-2\tilde{\Phi}}F_{\mu\nu}F^{\mu\nu}-2\p_{\mu}\tilde{\Phi}\p^{\mu}\tilde{\Phi}\right.\\
			\left. -\frac{1}{12}e^{-4\tilde{\Phi}}H_{k\mu\nu}H^{k\mu\nu}\right),
		\end{split}
	\label{actionST}
	\end{equation}
where $\tilde{\Phi}$ is the dilaton field and  $H_{k\mu\nu}$ is a third-rank tensor field defined as
	\begin{align}
		H_{k\mu\nu}\equiv&\p_{k}B_{\mu\nu}+\p_{\nu}B_{k\mu}+\p_{\mu}B_{\nu k}\nonumber\\
		&-2\left(A_{k}F_{\mu\nu}+A_{\nu}F_{k\mu}+A_{\mu}F_{\nu k}\right),
	\end{align}
with $B_{\mu\nu}$ being a second-rank antisymmetric tensor gauge field, known as Kalb-Ramond field. 

Written in Boyer-Lindquist coordinates $(t,r,\theta, \varphi)$,  the line element of the KS solution is:
	\begin{align}
		ds^{2}=&\left(1-\frac{2 M r}{\rho^{2}}\right)dt^{2}-\rho^{2}\left(\frac{dr^{2}}{\Delta}+d\theta^{2}\right)\nonumber\\
		&+\frac{4Mra\sin^{2}\theta}{\rho^{2}}dtd\varphi\nonumber\\
		&-\left[r\left(r+2 d\right)+a^{2}+\frac{2 M r a^{2}\sin^{2}\theta}{\rho^{2}}\right]\sin^{2}\theta\,d\varphi^{2},
		\label{ds2KS}
	\end{align}
where
	\begin{align}
		&d\equiv \frac{q^{2}_{\sss KS}}{2M},\\
		&\Delta\equiv r\left(r+2d\right)-2 M r+a^{2},\\
		&\rho^{2}\equiv r\left(r+2d\right)+a^{2}\cos^{2}\theta.
	\end{align}
As can be seen from the action \eqref{actionST}, there is a coupling between gravity and other fields, such as the dilaton field and gauge fields. The solutions for the non-gravitational fields of this spacetime are:
	\begin{align}
		&\tilde{\Phi}=-\frac{1}{2}\ln\frac{\rho^{2}}{r^2+a^2\cos^2\theta},\\
		&A_\mu dx^{\mu}=\frac{q_{\sss KS}}{\rho^2}r(dt-a\sin^{2}\theta d\varphi),\\
		&B_{t\varphi}=\frac{dra\sin^{2}\theta}{\rho^2}.
	\end{align}
	
The parameters $(M, q_{\sss KS}, a)$ are the mass, the electric charge and the angular momentum per unity mass of the KS black hole, respectively. The event horizon is localized at
	\begin{equation}
		r_{h_{\sss KS}} \equiv M-d+\sqrt{\left(M-d\right)^2-a^2}. 
		\label{rH_KS}
	\end{equation}

In the context of GR theory, the black hole that we shall use to compare our results is the charged rotating KN black hole. In Boyer--Lindquist coordinates, its line element reads:
		\begin{align}
	ds^{2}=&\left(1-\frac{2 M r - q_{\sss KN}^{2}}{\rho^{2}_{KN}}\right)dt^{2}-\rho^{2}_{KN}\left(\frac{dr^{2}}{\Delta_{KN}}+d\theta^{2}\right)\nonumber\\
	&+\frac{4 M a r\sin^{2}\theta-2a q_{\sss KN}^{2}\sin^{2}\theta}{\rho^{2}_{KN}}dtd\varphi\nonumber\\
	&-\frac{\left[(r^{2}+a^{2})^{2}-\Delta_{KN} a^{2}\sin^{2}\theta\right]\sin^{2}\theta}{\rho^{2}_{KN}}d\varphi^{2} .
	\label{ds2KN}
	\end{align}	
	The only non-gravitational field is associated with the electromagnetic vector potential, given by 
	\begin{equation}
	A_{\mu}dx^{\mu}=\frac{q_{\sss KN}}{\rho^{2}_{KN}}r(dt-a \sin^{2}\theta d\varphi), 
	\label{A_KN}
	\end{equation}		
	where
	\begin{align}
	&\Delta_{KN}\equiv r^{2}-2 M r+a^{2}+q_{\sss KN}^{2},\\
	&\rho^{2}_{KN}\equiv r^{2}+a^{2}\cos^{2}\theta.
	\end{align}
The KN black hole presents an event horizon at:
	\begin{equation}
	r_{h_{\sss KN}} \equiv M +\sqrt{M^{2}-a^{2}-q_{\sss KN}^{2}}.
	\end{equation}
	
	Both spacetimes have very similar geometrical aspects, but also present some distinctive features. For instance, the KS solutions allow a larger range of values of electric charge. Therefore, in order to compare our results, we shall use a normalized charge 
	\begin{equation}
	\Q_{(i)}=q_{(i)}/q^{ext}_{(i)},
\end{equation}	
 where $q^{ext}_{(i)}$ is the value of charge of an extreme black hole, with the subscript $(i)=KS,KN$ indicating the corresponding spacetime.
	\color{black}
\section{Scalar field in the KS background} \label{sec:2}
 
The dynamics of a massless scalar field $\Psi (x^{\mu})$ in KS spacetime is described by the Klein Gordon equation, which can be written in a covariant form as: 
	\begin{equation}
		\frac{1}{\sqrt{-g}}\p_{\mu}(\sqrt{-g}g^{\mu\nu}\p_{\nu}\Psi)=0,
		\label{KGeq}
	\end{equation}
	where $g$ is the determinant of the KS metric, whose contravariant components are $g^{\mu\nu}$.
	
In order to solve this equation, one can assume a separation form of $\Psi$ in terms of wavelike solutions, as given by:
	\small
		\begin{equation}
			\Psi=\sum_{l=0}^{+\infty}\sum_{m=-l}^{+l}\frac{U_{\omega l m}(r)}{\sqrt{r(r+2d)+a^{2}}}S_{\omega l m}(\theta)e^{i(m\varphi-\omega t)}.
			\label{Psi}
		\end{equation}
		\normalsize
The functions $S_{\omega l m}(\theta)$  are the oblate spheroidal harmonics,~\footnote{The KS solution belongs to a class of axisymmetric and asymptotically flat spacetimes that allow the separation of variables of the Klein-Gordon equation solutions as in Eq.~(\ref{Psi}), with the angular part being given in terms of the spheroidal harmonics \cite{Konoplya_etal:2018}.} that satisfy the following equation \cite{Abramovitz}:
	\begin{align}
		&\left(\frac{d^{2}}{d\theta^{2}}+\cot\theta\frac{d}{d\theta}\right)S_{\omega lm}\nonumber\\
		&+\left(\lambda_{lm}+a^{2}\omega^{2}\cos^{2}\theta-\frac{m^{2}}{\sin^{2}\theta}\right)S_{\omega lm}=0,
	\end{align}
where $\lambda_{lm}$ are the eigenvalues of the spheroidal harmonics. 

The radial part of the solution \eqref{Psi} satisfies a lengthy equation, which can be reshaped in terms of the tortoise coordinate $r_{\star}$, defined, in the KS spacetime, as
	\begin{equation}
		r_{\star}\equiv \int dr\left(\frac{r(r+2d)+a^{2}}{\Delta}\right).
	\end{equation}
With the aid of the tortoise coordinate, it is possible to rewrite the radial equation in a simpler form:
	
	\begin{equation}
		\left(\frac{d^{2}}{dr^{2}_{\star}}+V_{\omega lm}\right)U_{\omega lm}(r_{\star})=0,
		\label{RWeq}
	\end{equation}
where the function $V_{\omega lm}$ is given by:
	\begin{widetext}
	\begin{align}
		V_{\omega lm}=&-\frac{\Delta  \left(a^2 \left(-d^2-2 d M+2 d r-4 M r+r^2\right)+a^4+r \left(d^2 (2 M-r)-2 d^3+2 d M r+2 M r^2\right)\right)}{\left(a^2+2 d r+r^2\right)^4}\nonumber\\
		&-(\lambda _{lm} +a^2 \omega^2 -2 m a\omega)\frac{\Delta }{\left(a^2+2 d r+r^2\right)^2}+\frac{\left(a^2 \omega -a m+2 d r \omega +r^2 \omega \right)^2}{\left(a^2+2 d r+r^2\right)^2}.
		\label{V_KS}
	\end{align}\\
\end{widetext}
			
\begin{figure}
		\includegraphics[width=7cm]{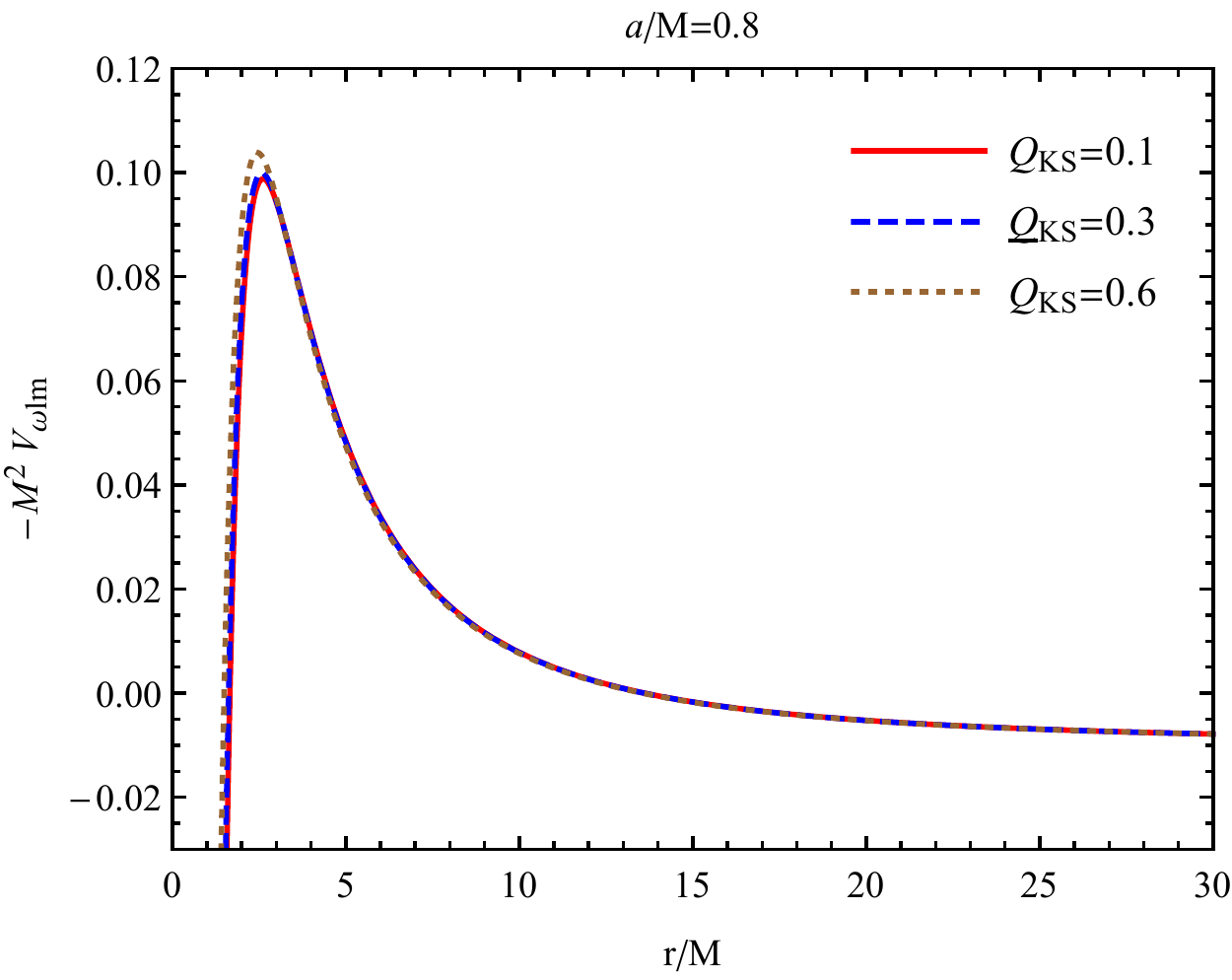}\hspace{0.2cm}
		\includegraphics[width=7cm]{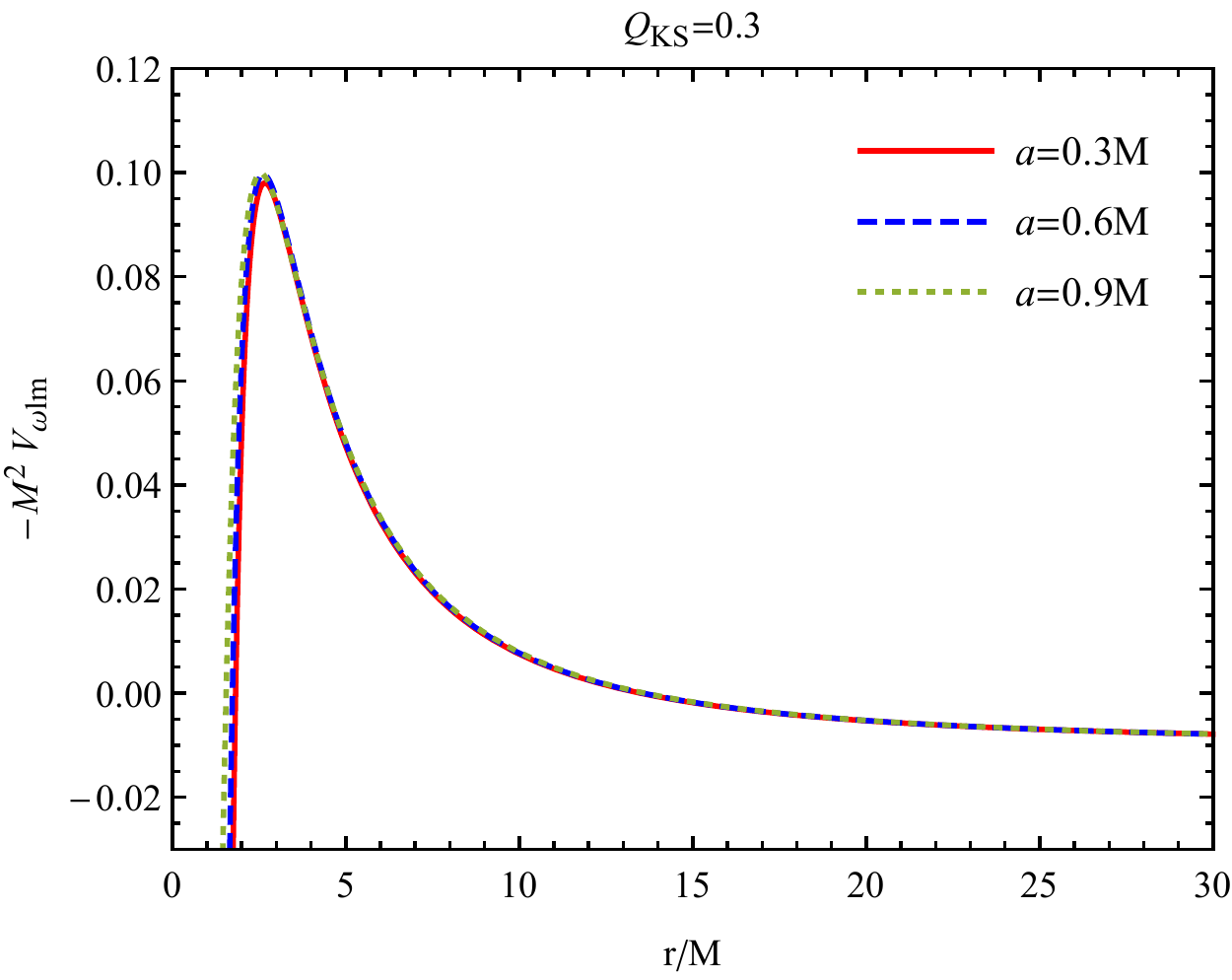}
		\caption{$V_{\omega lm}$ as a function of $r$, for different values of the KS black hole normalized charge (top panel) and different values of the KS black hole angular momentum (bottom panel). For both panels we have chosen $l=m=1$ and $M\omega=0.1.$}
		\label{Fig1}
\end{figure}

In Fig~\ref{Fig1}, we plot $V_{\omega lm}$ for different values of the normalized charge $Q_{\sss KS}$ (top panel) and rotation parameter $a$ (bottom panel), setting $M\omega = 0.1$ and $l=m=1$ in both cases. One can notice that $V_{\omega lm}$ is larger for higher values of the BH charge and spin. In the asymptotic limit, $V_{\omega lm}$ tends to $\omega^{2}$, as in the Kerr-Newman case \cite{L_C_C:2017}.
	
	When dealing with absorption and scattering phenomena, it is common to use a set of independent solutions called {\it in} modes, which correspond to incoming waves from the past null infinity, part of which is transmitted to the event horizon while the remaining part is reflected to the future null infinity. In the asymptotic limits (infinity and event horizon), these modes have the following behavior:
	\begin{equation}
		U_{\omega lm}(r_{\star})~\begin{cases}
		\mathcal{I}_{\omega lm}U_{I}+\mathcal{R}_{\omega lm}U^{*}_{I} \hspace{0.3cm} (r_{\star}/M\rightarrow +\infty),\\
		\mathcal{T}_{\omega lm}U_{T} \hspace{1.9cm} (r_{\star}/M\rightarrow -\infty),
		\end{cases}
	\end{equation}
with the symbol $*$ denoting complex conjugation. The terms $U_{I}$ and $U_{T}$ can be expanded as:
	\begin{align}
		&U_{I}=e^{-i\omega r_{\star}}\sum_{j=0}^{N}\frac{h_{j}}{r^{j}},\\
		&U_{T}=e^{-i(\omega-m\Omega_{H})r_{\star}}\sum_{j=0}^{N}g_{j}(r-r_{h_{\sss KS}})^{j},
	\end{align}
where 
	\begin{equation}
		\Omega_{H}=\frac{a}{2 Mr_{h_{\sss KS}}}
	\end{equation}
is the angular velocity of the event horizon.
The  coefficients $h_{j}$ and $g_{j}$ are constants found by solving Eq.~\eqref{RWeq}. The terms $\mathcal{I}$, $\mathcal{R}$ and $\mathcal{T}$ are related to the incident, reflected and transmitted parts of the wave, respectively. It can be show that they obey the relation:
	\begin{equation}
		\left|\frac{\mathcal{R}_{\omega lm}}{\mathcal{I}_{\omega lm}}\right|^{2}=1-\frac{\omega-m\Omega_{H}}{\omega}\left|\frac{\mathcal{T}_{\omega lm}}{\mathcal{I}_{\omega lm}}\right|^{2}.
	\end{equation}
	
For $\omega< m\Omega$ the reflection coefficient is greater than 1, corresponding to the superradiance phenomena. The equality $\omega= m\Omega$ also corresponds to the minimum value of frequency for which the black hole can be overspun into a naked singularity \cite{Duztas:2018adf}.

\section{Absorption cross section} \label{sec:3}
Considering a scalar massless plane wave  whose incidence angle is $\gamma$, one can  find the partial absorption cross section through partial wave methods \cite{Futterman_etal:1988}, as:
	\begin{equation}
		\sigma_{lm}=\frac{4\pi^{2}}{\omega^{2}}|S_{\omega lm}(\gamma)|^{2}\left(1-\left|\frac{\mathcal{R}_{\omega lm}}{\mathcal{I}_{\omega lm}}\right|^{2}\right).
		\label{partialabs}
	\end{equation}
The total absorption cross section is obtained as a sum of the partial contributions $\sigma_{lm}$ over all allowed values of $l$ and $m$:
	\begin{equation}
		\sigma=\sum_{l=0}^{\infty}\sum_{m=-l}^{+l}\sigma_{lm}.
	\end{equation}
	
It is possible to decompose the total absorption cross section in a sum of corotating modes, given by
	\begin{equation}
		\sigma^{co}=\sum_{l=1}^{\infty}\sum_{m=1}^{l}\sigma_{lm},
	\end{equation}
and counterrotating modes, given by
	\begin{equation}
		\sigma^{counter}=\sum_{l=1}^{\infty}\left(\sigma_{l0}+\sum_{m=1}^{l}\sigma_{l(-m)}\right).
	\end{equation}
\vspace{0.1cm}

\subsection{Low-frequency regime}
There are some analytical approximations in the literature for the absorption cross sections that allow us to test the accuracy of our numerical results. For the low-frequency regime, Higuchi found in 2001 \cite{Higuchi:2001} that the absorption cross section of a stationary and circularly symmetric black hole with any dimensions tends to the area of the event horizon. 

The event horizon area of the KS black hole is given by
\begin{equation}
A_{h}=\int_{0}^{\pi}\int_{-\pi}^{\pi}\sqrt{g_{\theta\theta}g_{\varphi\varphi}}d\varphi d\theta=4\pi\left(r_{h_{\sss KS}}^{2}+a^{2}+2dr_{h_{\sss KS}}\right).
\end{equation}
As we shall see in Sec.~\ref{sec:4}, our numerical results are in excellent agreement  with the corresponding low-frequency approximation.

\subsection{High-frequency regime}

In the high-frequency regime, the absorption cross section is intimately related with the properties of the photon orbits. This relation is very well represented by the so-called sinc approximation. Firstly proposed by Sanchez \cite{Sanchez:1978}, for the Schwarzschild black hole, this approximation was later derived using complex angular momentum techniques~\cite{Decanini_etal:2011} (c. f. also \cite{Caio_Crispino:2013,Carol_Leite_Crispino:2018}).
One can show that:
\begin{equation}
	\sigma\approx\sigma_{geo}\left[1-\frac{8\pi\beta e^{-\pi\beta}}{\Omega_N^{2}b^{2}_{c}}\text{sinc}\left(\frac{2\pi\omega}{\Omega_N}\right)\right],
	\label{sincapp}
\end{equation}\\
where $\beta\equiv\Lambda/\Omega_N$, with $\Lambda$ being the Lyapunov exponent, $\Omega_N$ the orbital frequency for the null orbit, $b_{c}$ the critical impact parameter, and $\sigma_{geo}\equiv\pi b_{c}^{2}$. For the on-axis case of the KS black hole, we have: 
\begin{widetext}
	\begin{equation}
		r_{c}=M-d+2\sqrt{M^{2}-\left(a^{2}+4Md+2d^{2}\right)/3}\cos\left[\frac{1}{3}\arccos\left[\frac{M\left((M-d)^{2}-a^{2}\right)}{\left(M^{2}-\left(a^{2}+4Md+2d^{2}\right)/3\right)^{3/2}}\right]\right],
	\end{equation}
	\begin{equation}
		b_{c}^{2}=\frac{2(r_{c}+d)\left[r_{c}(r_{c}+2d)+a^{2}\right]}{r_{c}+d-M},
	\end{equation}
	\begin{equation}
		\Lambda=\frac{4 \sqrt{2 a^2-b_{c}^2+4 d^2+12 d r_{c}+6 r^2}}{T_{0}\sqrt{\mathcal{K} }}K\left(-\frac{a^2}{\mathcal{K} }\right),	
	\end{equation}
	\begin{equation}
		T_{0}=\frac{4}{\sqrt{\mathcal{K}}}\left[\mathcal{K} E\left(-\frac{a^{2}}{\mathcal{K}}\right)+\left(\frac{(r_{c}(r_{c}+2d)+a^{2})^{2}}{\Delta}-(a^{2}+\mathcal{K})\right)K\left(-\frac{a^{2}}{\mathcal{K}}\right)\right],
	\end{equation}
\end{widetext}
where $K(k)$ and $E(k)$ are the complete elliptic integrals of the first and second kind, respectively, $\mathcal{K}\equiv \mathcal{C}-(a-\mathcal{L}_{z})^{2}$ is a Carter-like constant, $\mathcal{L}_{z}\equiv L_{z}/E$ is the azimuthal angular momentum per unit energy, $\mathcal{C}$ is a constant of motion, $r_{c}$ is the critical radius and $T_{0}$ is the latitudinal period.

As we shall see in Sec.~\ref{sec:4}, our numerical results are in excellent agreement with the sinc approximation.

\color{black}
\section{Results}\label{sec:4}

In this section we show a selection of our results for the total absorption cross section of the KS black hole in the Einstein frame for different values of the incidence angle $\gamma$ of the scalar wave.

\begin{figure}
	\includegraphics[width=7cm]{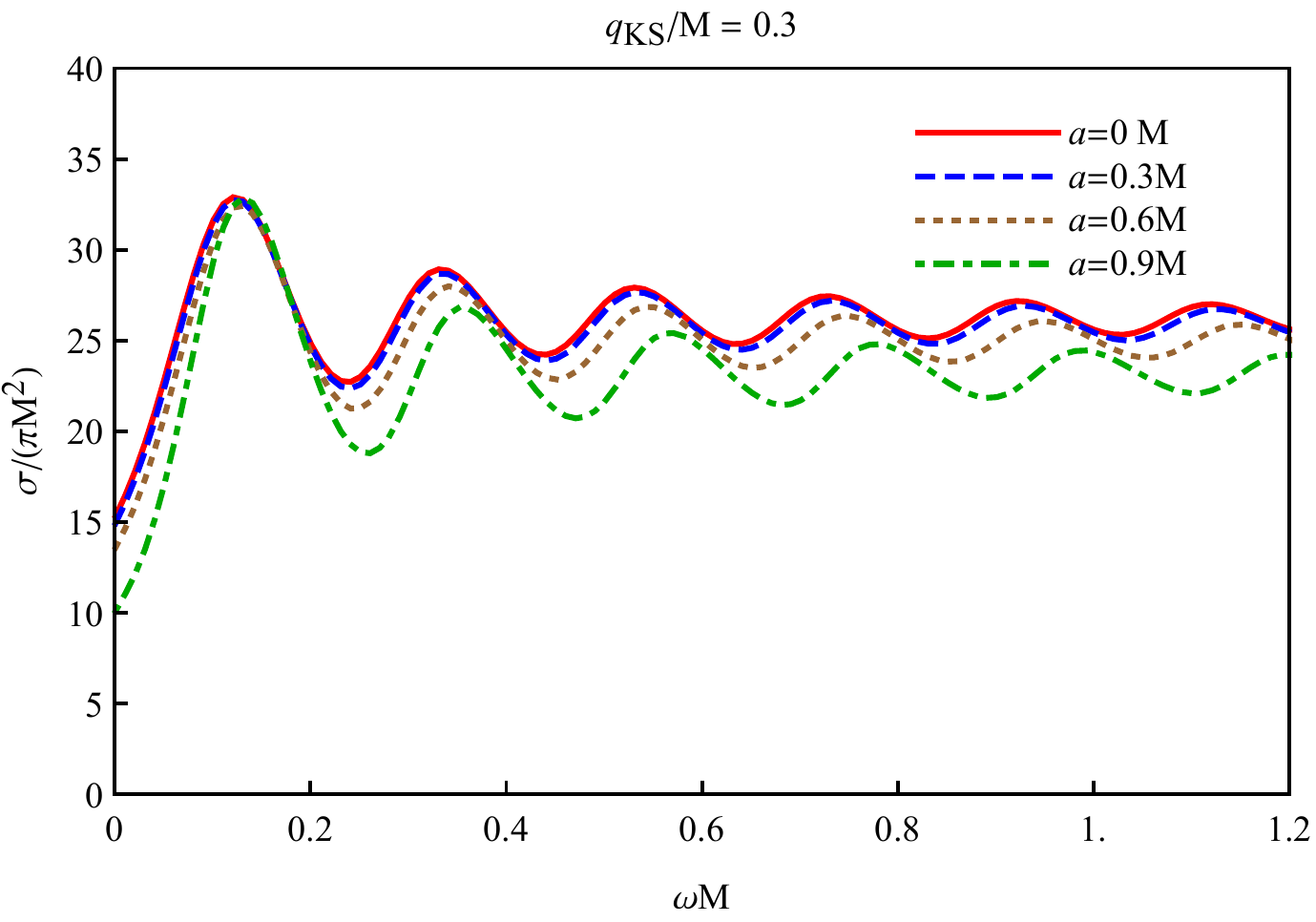}\hspace{0.2cm}
	\includegraphics[width=7cm]{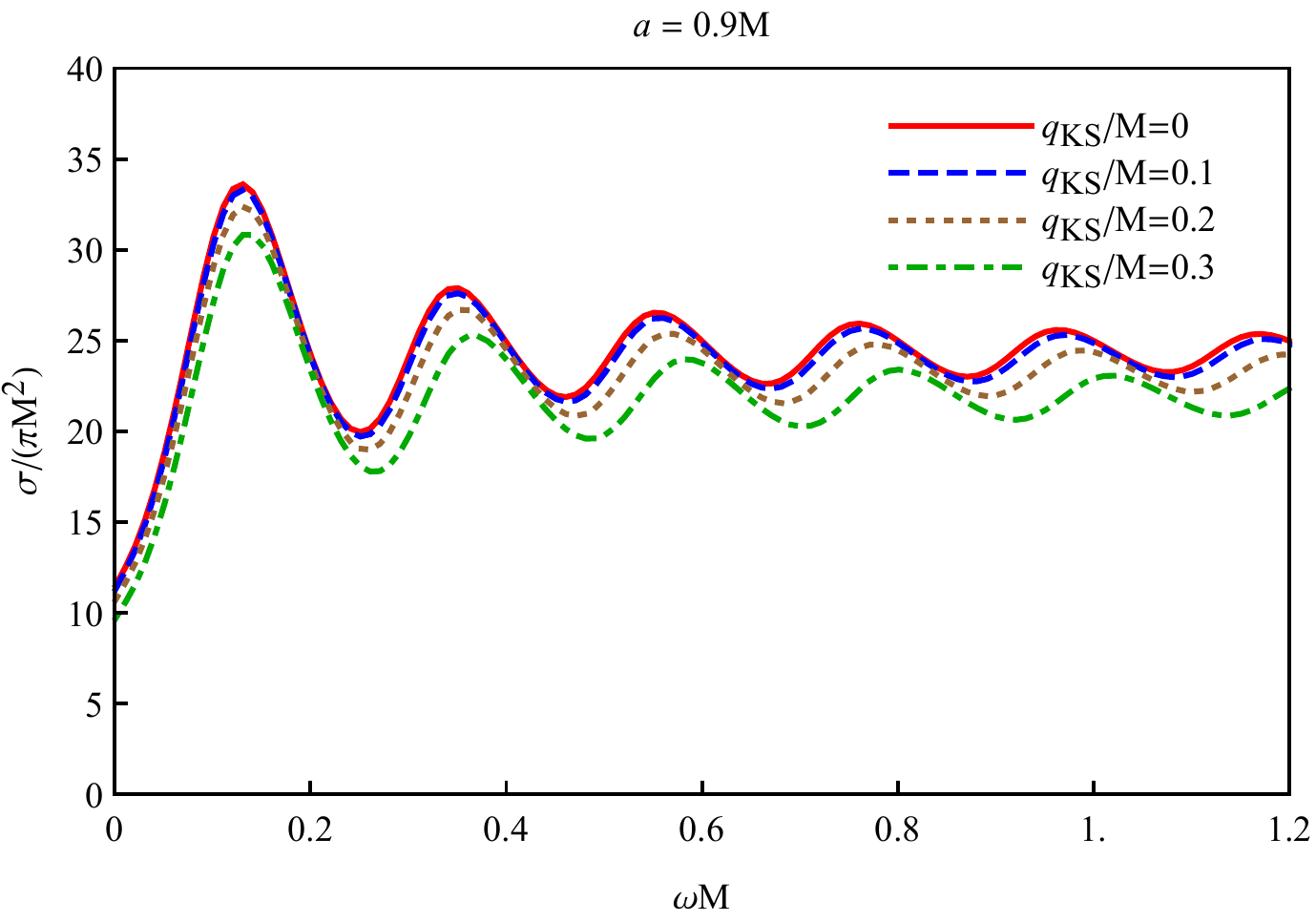}
	\caption{Total absorption cross section for fixed charge (top) and angular momentum per unity mass (bottom) of the KS black hole. Both panels are plotted for polar incidence ($\gamma=0$).}
	\label{Fig2}
\end{figure}

In Fig.~\ref{Fig2}, we exhibit the total absorption cross section for polar incidence ($\gamma=0$) with fixed values of charge (top panel) and fixed values of the rotation parameter (bottom panel). One can notice that as the values of charge and angular momentum increase, the total absorption cross section decreases. Furthermore, one can see a regular oscillatory  behavior around the high-frequency limit.

\begin{figure}
	\includegraphics[width=7cm]{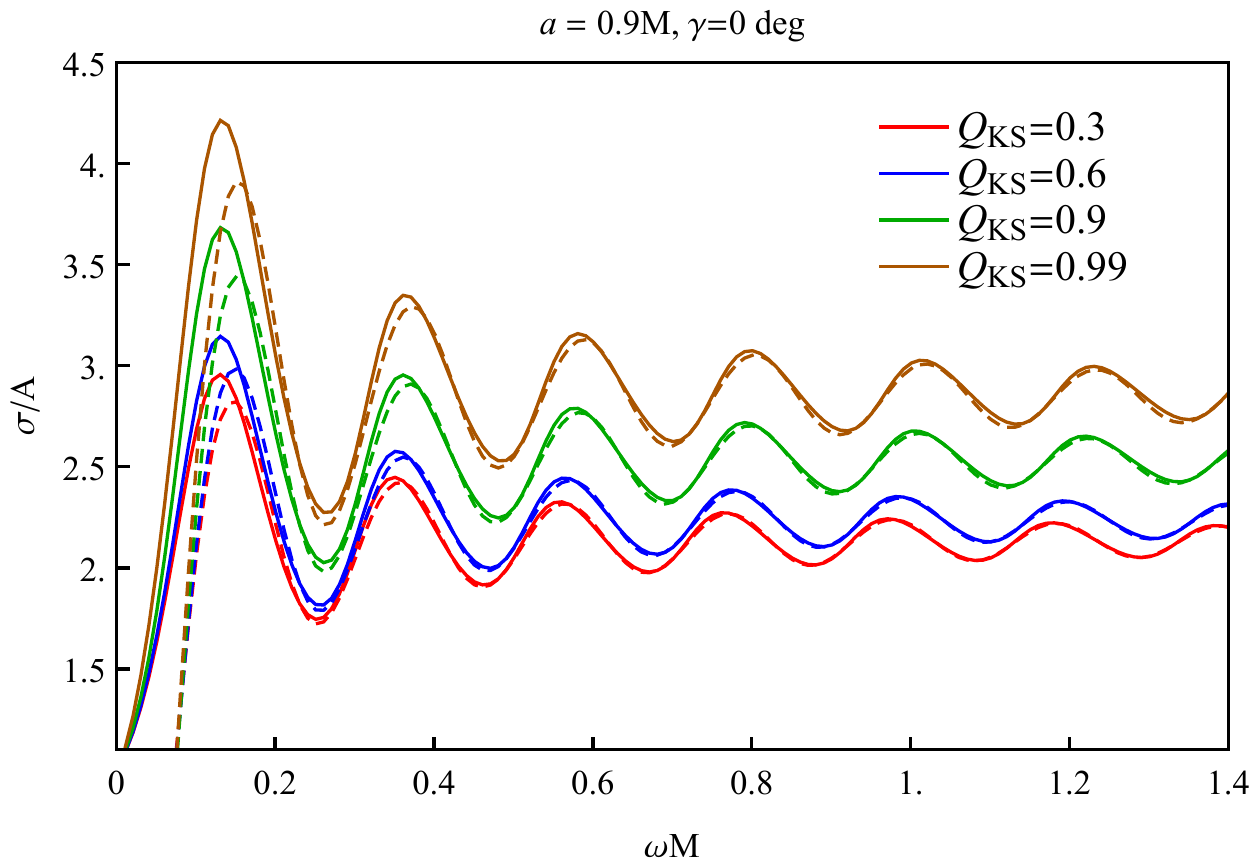}\hspace{0.2cm}
	\caption{Sinc approximation for the on-axis ($\gamma = 0$) case. The solid lines are the numerically computed absorption cross sections and the dashed lines represent the sinc approximation.}
	\label{Fig_sinc}
\end{figure}

In Fig.~\ref{Fig_sinc} we compare our numerical results for polar incidence
with the corresponding ones obtained via the sinc approximation, showing excellent agreement.

\begin{figure}
	\includegraphics[width=7cm]{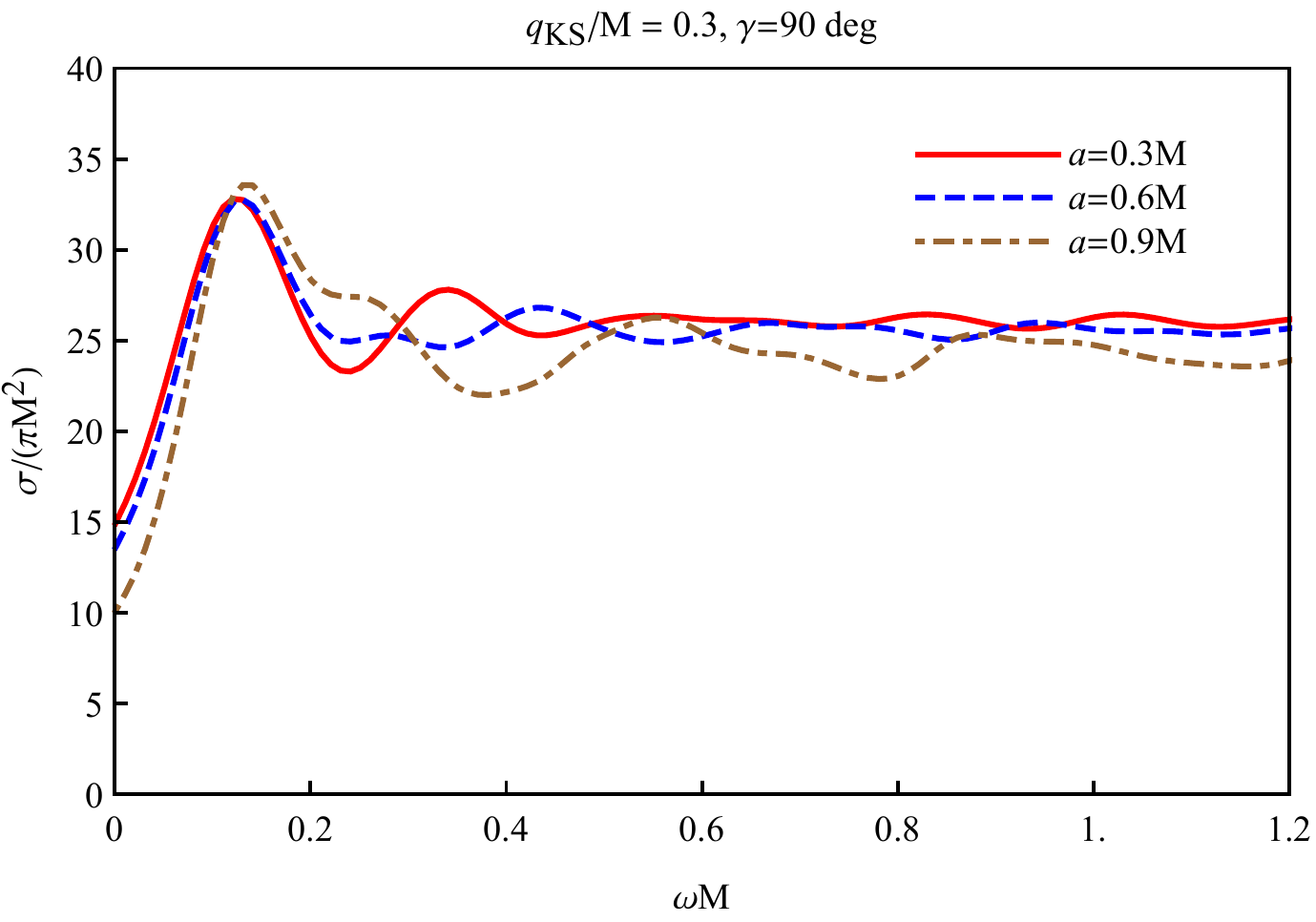}\hspace{0.2cm}
	\includegraphics[width=7cm]{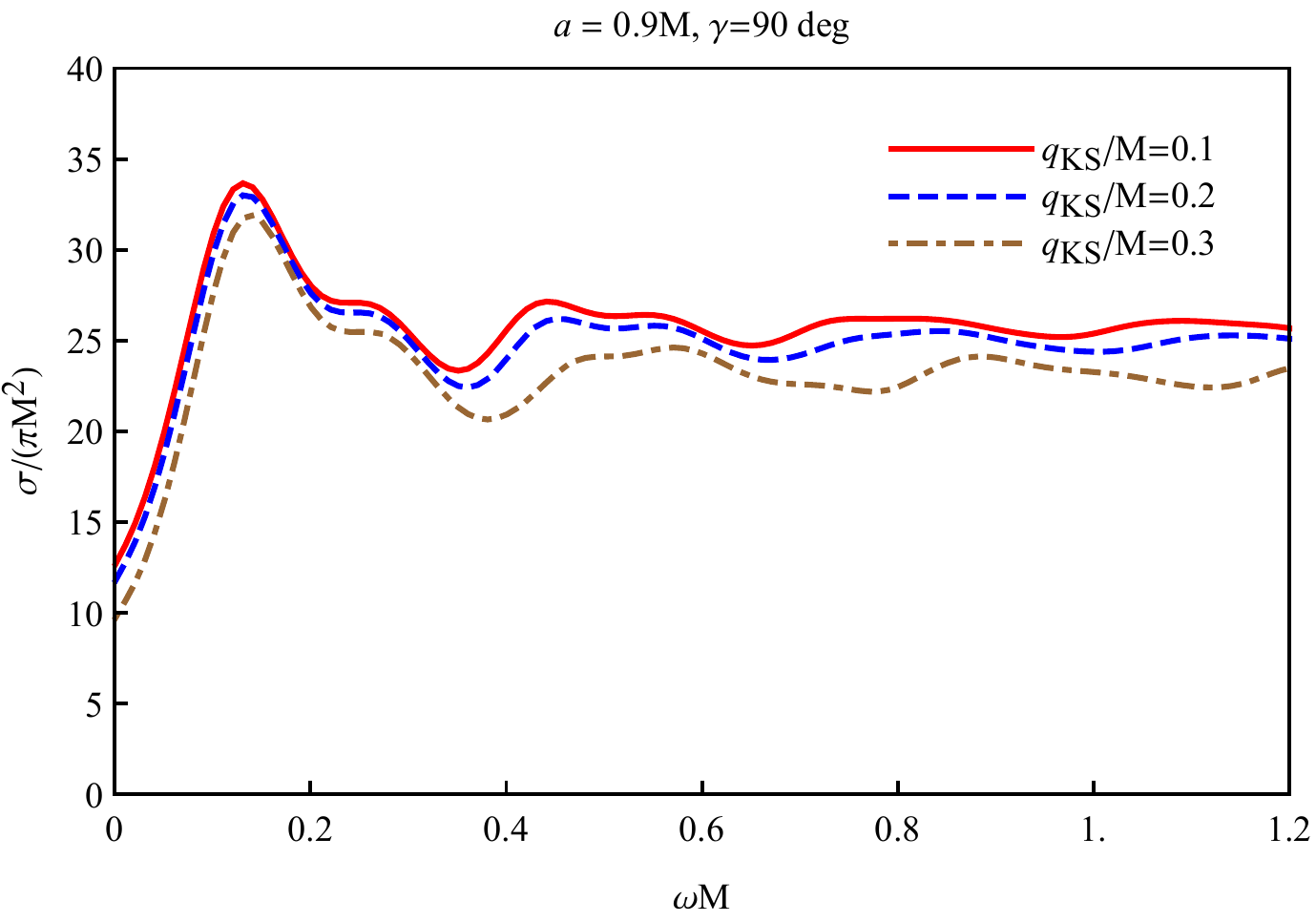}
	\caption{Total absorption cross section for different values of the rotation parameter (top) and electric charge (bottom), plotted for equatorial incidence.  }
	\label{Fig3}
\end{figure}

Illustrated by Fig.~\ref{Fig3}, the equatorial incidence case ($\gamma = 90$ deg) reveals a distinctive behavior when compared with in the on-axis case. One can identify  an oscillatory pattern less regular than the one  exhibited in Fig.~\ref{Fig2}. On the top panel the total absorption cross section is plotted for fixed value of the KS black hole charge ($q_{\sss KS}/M=0.3$) and $a/M=0.3$, $0.6$ and $0.9$, while in the bottom panel we consider KS black holes with same angular momentum ($a/M=0.9$) and different values of the charge, namely $q_{\sss KS}/M=0.1$, $0.3$ and $0.5$. Note that although the  increase of spin and charge leads generally to a smaller absorption cross section, the plots in the bottom panel preserves its essential shape, what does not occur in the top panel.

\begin{figure}
		\includegraphics[width=8cm]{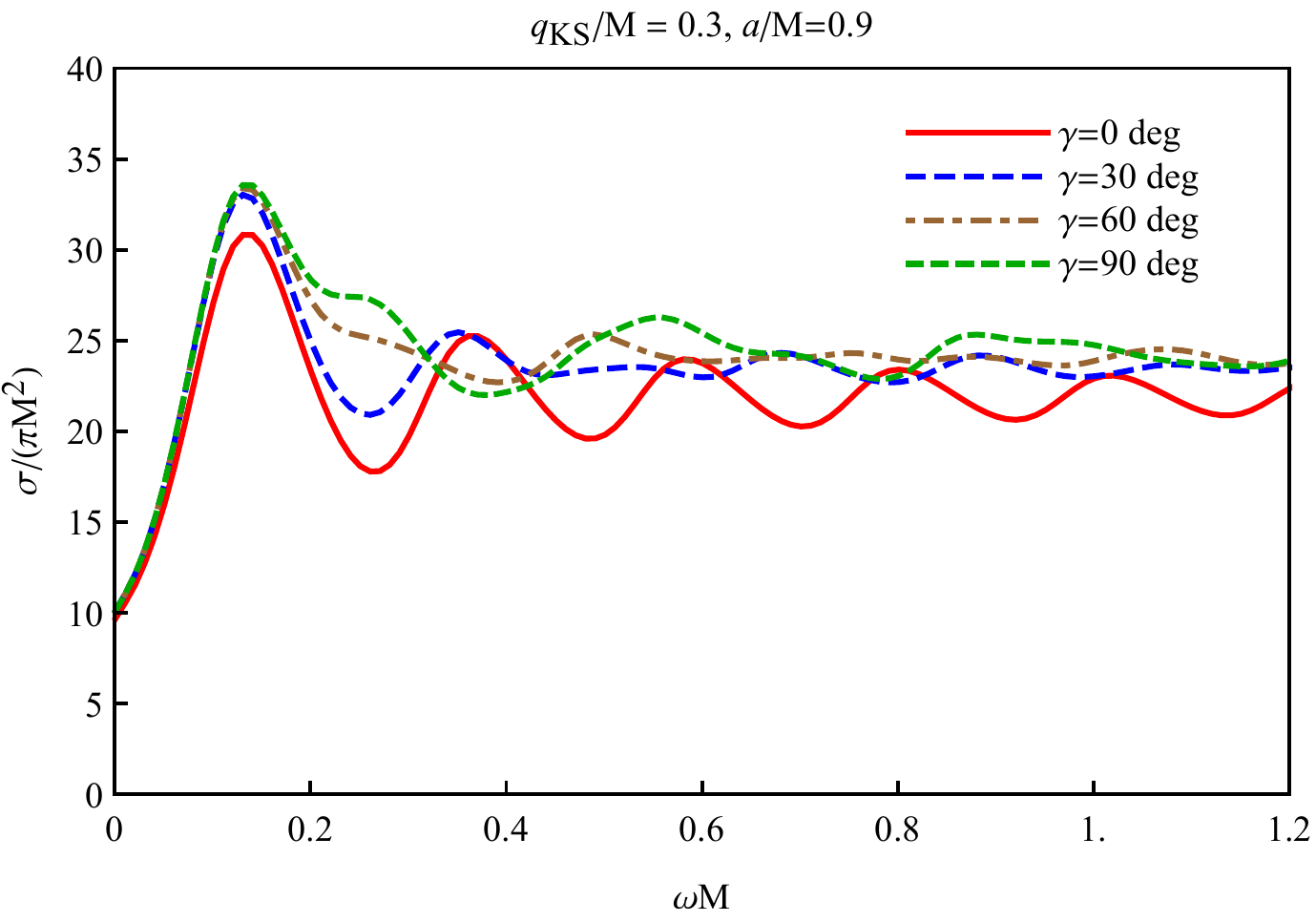}\hspace{0.2cm}
		\caption{Total absorption cross section of a KS black hole with $q_{\sss KS}/M=0.3$ and $a/M=0.9$ for different values of the incidence angle.}
		\label{Fig6}
\end{figure}

In Fig.~\ref{Fig6} we show the total absorption cross section for different choices of the scalar wave incidence angle. One can note that as we move away from the on-axis case ($\gamma=0$ deg), the  behavior of the absorption cross section is less regular.

\begin{figure}
 	\includegraphics[width=7cm]{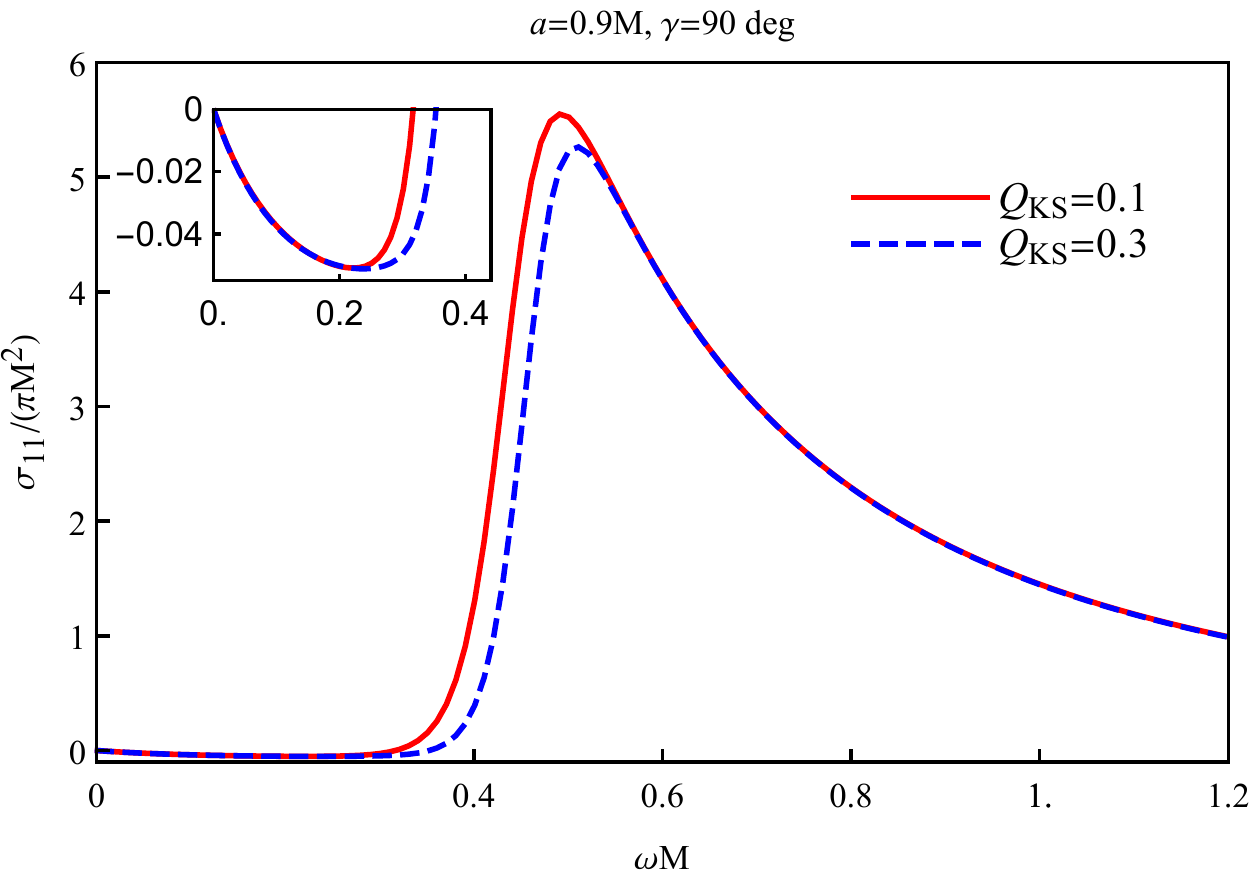}\hspace{0.2cm}
 	\includegraphics[width=7cm]{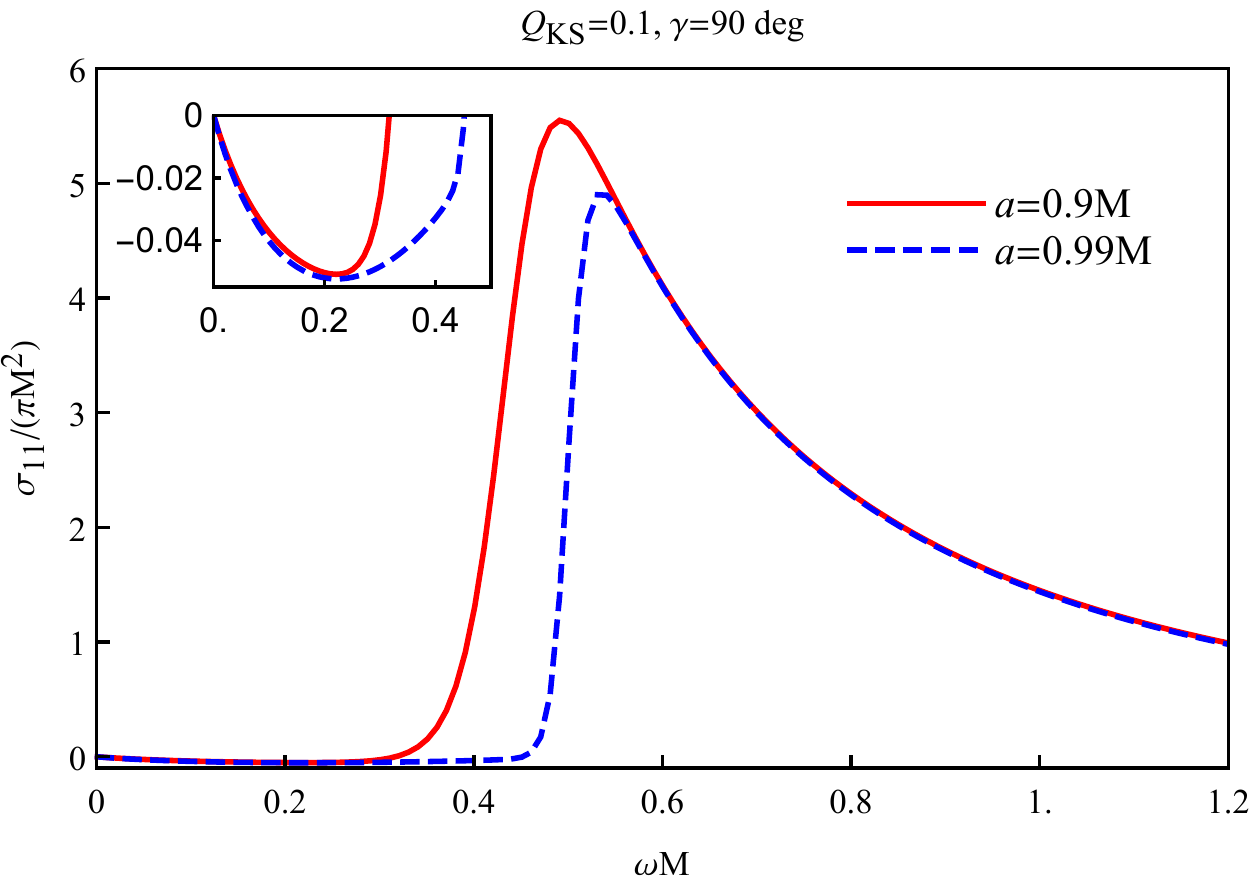}
 	\caption{Partial absorption cross section with $l=m=1$, for fixed angular momentum per unity mass (top) and normalised charge (bottom). Both panels are plotted for equatorial incidence ($\gamma=90$ deg). The insets emphasize the occurrence of superradiance.}
 	\label{Fig7}
 \end{figure}

Superradiance is associated to the amplification of the reflected waves, what in this context is characterized by a negative partial absorption cross section. In Fig.~\ref{Fig7} we plot the partial absorption cross section for $l=m=1$, highlighting the negative values in the insets as a clear manifestation of the superradiance phenomenon.

\begin{figure}
	\includegraphics[width=7cm]{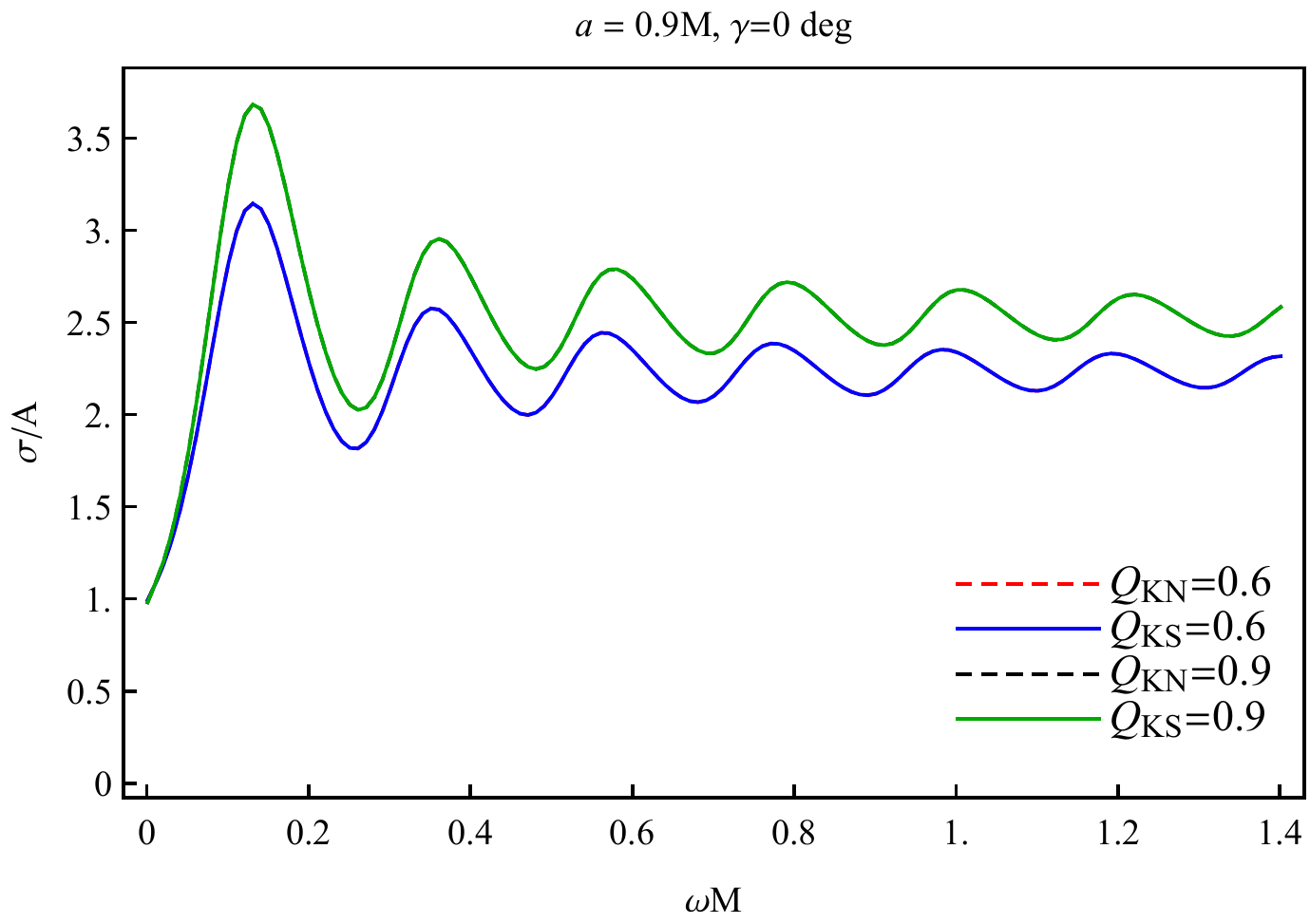}\hspace{0.2cm}
	\includegraphics[width=7cm]{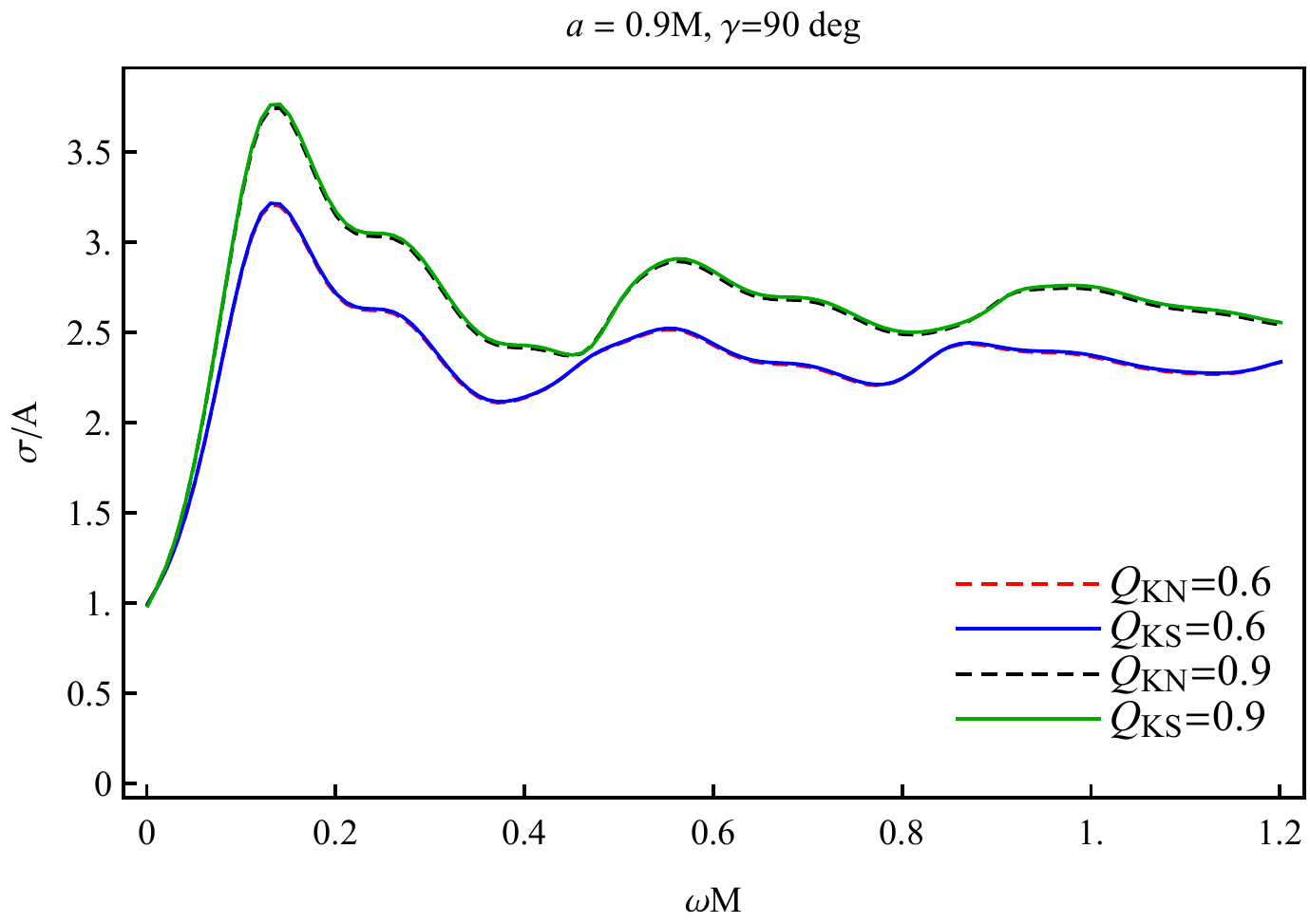}
	\caption{Comparison between the ratio of absorption cross section and area of the black holes as a function of the normalised charge $\Q_{(i)}=q_{(i)}/q^{ext}_{(i)}$. The top panel corresponds to a wave with polar incidence ($\gamma=0$ deg) whereas the bottom panel corresponds to a wave with equatorial incidence ($\gamma=90$ deg).}
	\label{Fig8}
\end{figure}

We plot the absorption cross section divided by the horizon area of KN and KS black holes in Fig.~\ref{Fig8}, for the on-axis ($\gamma=0$ deg) and equatorial incidence ($\gamma=90$ deg) cases. 
We see that, in general, the KS black hole has a larger total absorption cross section than the corresponding KN case.
This is in agreement with the fact that the function $V_{\omega lm}$ of the KS spacetime is generally smaller than the corresponding one of the KN spacetime.~\footnote{We notice that for some low-frequency values and black holes close to extremality, the KN potential can be smaller than the KS one.}
We also notice that in the low-frequency limit the absorption cross section agrees with the result of Ref.~\cite{Higuchi:2001}, i.e., tends to the area of the black hole horizon.

\begin{figure}
		\includegraphics[width=7cm]{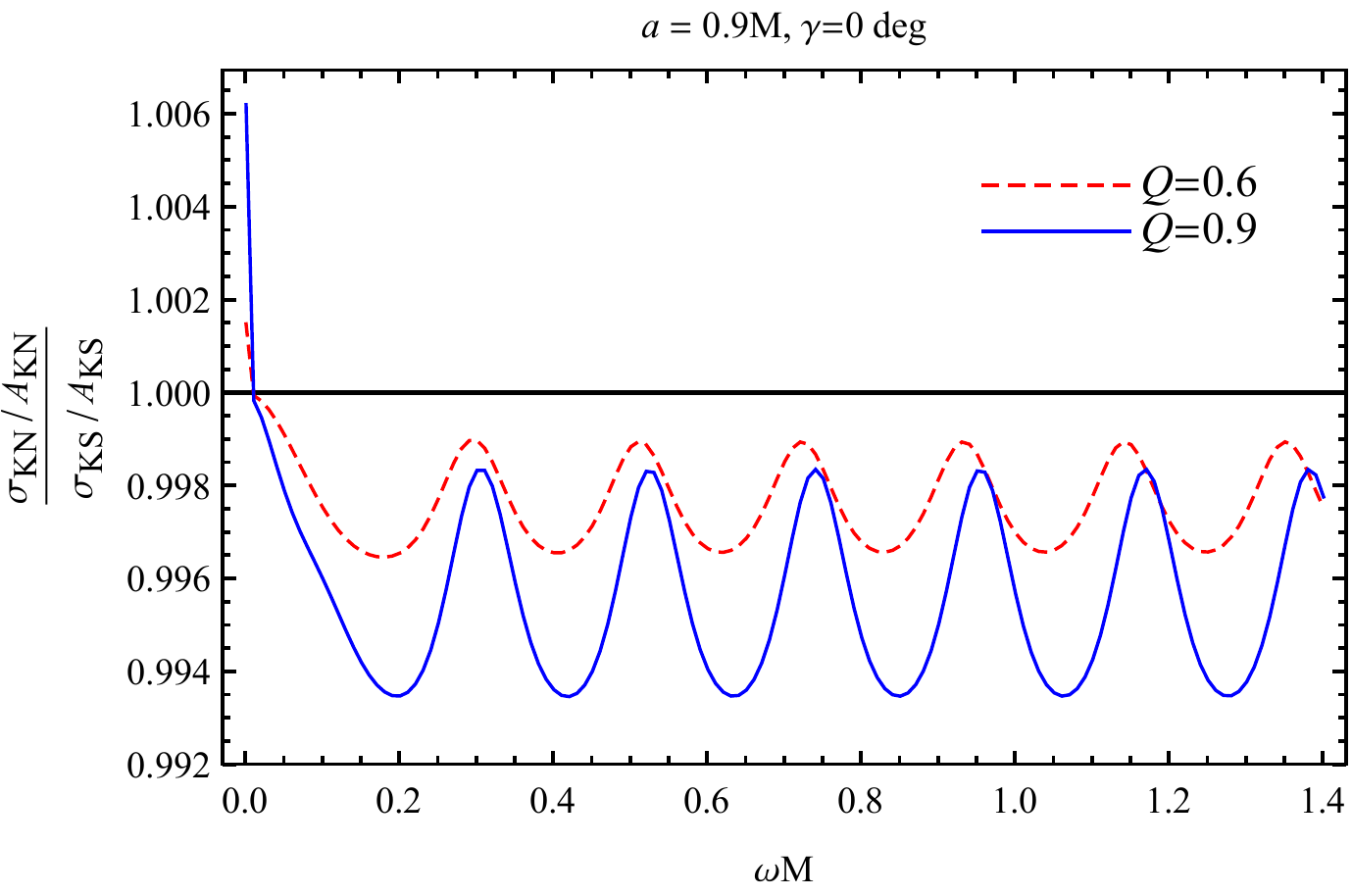}\hspace{0.2cm}
		\includegraphics[width=7cm]{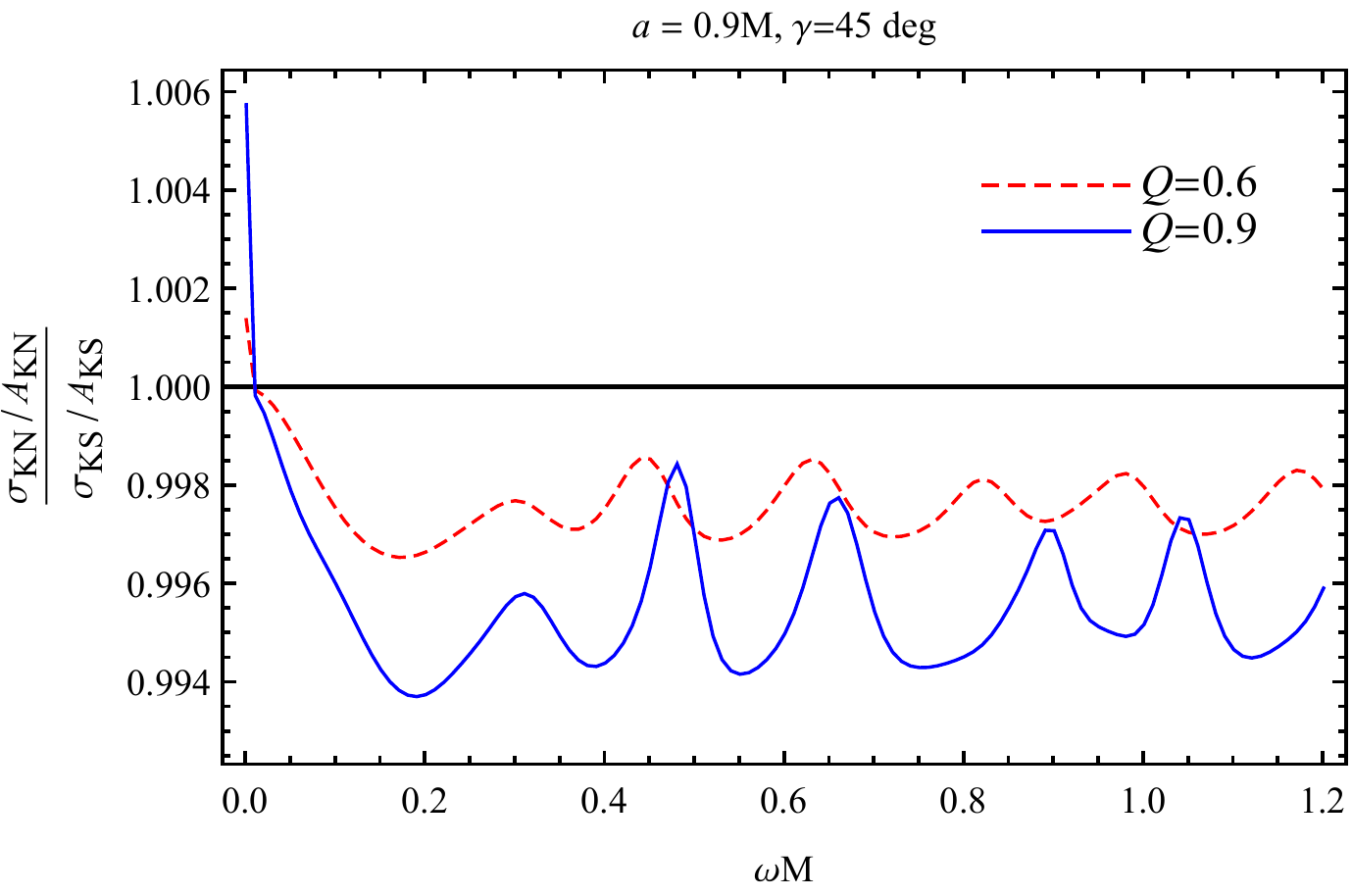}\hspace{0.2cm}
		\includegraphics[width=7cm]{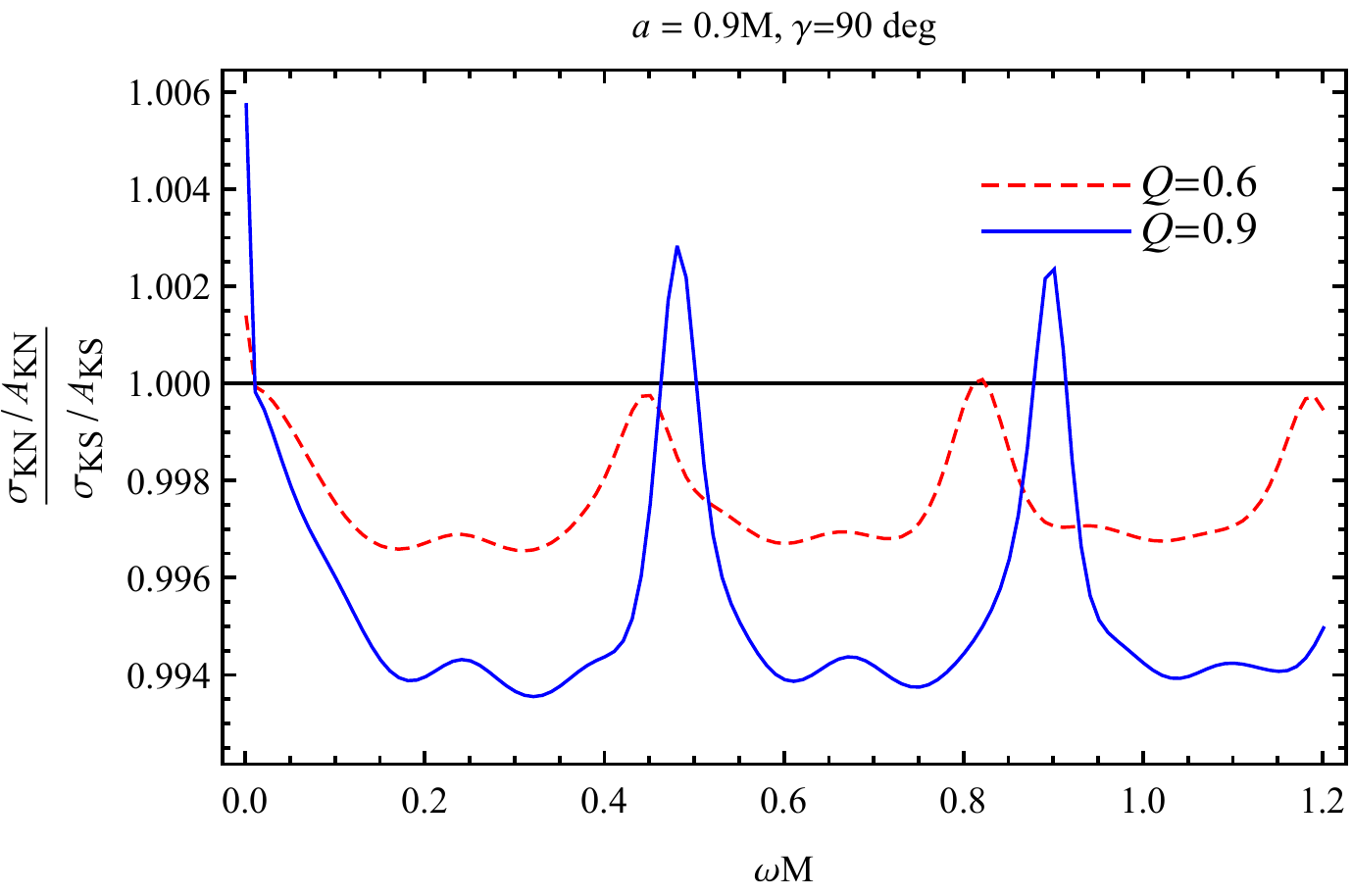}
		\caption{Ratio between the absorption cross section and the areas of the black holes for a wave with incidence angle $\gamma = 0$ deg (top), $\gamma = 45$ deg (middle) and $\gamma = 90$ deg (bottom). 
		}
		\label{Fig10}
\end{figure}

In Fig.~\ref{Fig10} we plot the ratio between the absorption cross section and the areas of the KS and KN black holes, namely $(\sigma_{KN} /A_{KN})/(\sigma_{KS} /A_{KS})$. We see that this quantity is very close to unity, implying that the difference between the absorption cross section of both black holes is very small. Furthermore, we see that, in general, this quantity is larger for KS black holes than for KN black holes.

\section{Final remarks}\label{sec:remarks}

We computed numerically the absorption cross section for KS black holes. We have shown that,  for the on-axis case ($\gamma=0$), there is a well behaved oscillatory pattern, resulting from the contributions of waves with different angular momentum, $l$. Besides that, the increase of the parameters of the black hole (charge and angular momentum), decreases the absorption cross section.
	
For the off-axis case, it can be noted that a less well behaved pattern is present. This  is a feature usually seen in stationary and axisymmetric spacetimes~\cite{Caio_Crispino:2013,L_C_C:2017}. It is explained by the distinct contributions coming from the co- and counterrotating modes. Moreover, the absorption cross section becomes more irregular for higher values of the black hole charge and angular momentum.

When the condition $\omega <m\Omega_H$ is obeyed, the partial absorption cross section is negative. This is due to the superradiant scattering, which implies in a reflection coefficient greater than 1. For the scalar case, this effect is more pronounced for the $l=m=1$ wave~\cite{BC2016}. Although the partial absorption cross section can be negative, the total absorption cross section remains positive, due to the fact that to compute the total absorption cross section we sum over both the positive and negative $m$ contributions. 

In  order to compare the absorption cross section of the KS black hole with the one of a KN black hole, we defined the normalized charge $\Q$. We have shown that both spacetimes have qualitatively similar behaviors of the massless absorption cross section, although the KS case generally 
presents a larger absorption cross section than the KN case, for the same values of the pair ($\omega M, \Q$).
The difference between the two cases increases as the black holes approach the extremality, albeit being very small.
A small difference between physical quantities associated to KN and KS black holes with the same normalized charge is also manifest in the case of their shadows~\cite{Hioki_Miyamoto:2008,Siahaan:2020,Xavier:2020egv,Cayuso:2019ieu}. 

The electromagnetic sector of the Kerr-Sen solution is more involved than the Kerr-Newman case, due to the interaction with additional fields present in the former case. It would be interesting to analyze the case of the incidence of an electrically charged scalar field, and the absorption cross section will depend on whether the electromagnetic interaction between the black hole and the field is attractive or repulsive.

\begin{acknowledgments}
The authors thank Funda\c{c}\~ao Amaz\^onia de Amparo a Estudos e Pesquisas (FAPESPA),  Conselho Nacional de Desenvolvimento Cient\'ifico e Tecnol\'ogico (CNPq) and Coordena\c{c}\~ao de Aperfei\c{c}oamento de Pessoal de N\'{\i}vel Superior (Capes) - Finance Code 001, in Brazil, for partial financial support. We are grateful to Carlos
Herdeiro for useful discussions.
This work has further been supported by  the  European  Union's  Horizon  2020  research  and  innovation  (RISE) programme H2020-MSCA-RISE-2017 Grant No.~FunFiCO-777740.
\end{acknowledgments}

	{}
\end{document}